# Experimental preparation of pure superposition states of atoms via elliptically polarized bichromatic radiation

Sergei A. Zibrov, V.L. Velichansky

*P.N. Lebedev Institute of Physics, Moscow, 117924, Russia,*
*Engineering and Physics State University, Moscow,115409, Russia*

A.V. Taichenachev, V.I. Yudin

*Institute of Laser Physics, Novosibirsk, 630090, Russia,*
*Novosibirsk State University, 630090, Russia*

A.S. Zibrov

*Physics Department, Harvard University, Cambridge, MA, 02138, USA*
*P.N. Lebedev Institute of Physics, Moscow, 117924, Russia,*

We propose a simple and effective way of creating pure dark superposition states. The generation of pure states is carried out by using bichromatic radiation with controllable polarization ellipticity. We derived analytic formulas for polarization elipticity to obtain pure dark states of different Zeeman sublevels of alkali atoms. Experimentally we accumulated ∼60% of the atoms in the 0-0 dark state of the $D_1$ line of $Rb^{87}$. © 2005 Optical Society of America

The pure superposition state is a basic concept of quantum physics, and plays an important role in fundamental studies and various practical applications. The coherent population trapping effect (CPT) is one of many possible ways for the preparation of a superposition state by optical means.[1] Using bichromatic laser radiation Λ atoms are collected in non-absorbing dark state, resulting in bleaching of the absorptive media at zero two-photon detuning. CPT intensively studied last decade for a wide range of applications. Among them are atom cooling, high sensitive magnetometery,[2] atomic clocks and masers,[3] and recently emerged, new robust techniques attractive for scalable quantum networks and computing.[4] In all these applications, the quality of the prepared pure states defines their ultimate limit. Some particular ways of preparation of the 0-0 pure superposition states were found in.[5–7] In[5] "push-pull" optical pumping technique with harmonic amplitude modulation of the applied bichromatic radiation was used, and in[6,7] the authors prepared pure superposition states by specific polarization of radiation.

Strong need for preparation of pure superposition states exists in quantum optics as well. A CPT-based quantum memory and tunable single-photon light source is demonstrated in.[9] Kuhn et al., used stimulated Raman scattering adiabatic passage (STIRAP) to make a single-photon source of radiation.[8] In all these cases the fidelity of the created pure state is crucial.

In this letter we experimentally realize a simple, effective way of preparing pure superposition states in the system of Zeeman substates with any equal magnetic quantum numbers of the ground-state hyperfine levels of alkali atoms. Our method is based on the relevant selection of the elliptical polarization of the components of the bichromatic radiation.

Consider the resonant interaction between atoms and a bichromatic field propagating along the $Z$ axis:

$$\mathbf{E}(z,t) = E_1(z)\mathbf{a_1}e^{-i\omega_1 t} + E_2(z)\mathbf{a_2}e^{-i\omega_2 t} + c.c. \qquad (1)$$

The frequency components of this field have arbitrary complex amplitudes $E_{1,2}$ and elliptical polarizations described by unit vectors $\mathbf{a_{1,2}}$. These vectors can be written in the spherical basis ($\mathbf{e_{\pm 1}} = \mp(\mathbf{e_x} \pm i\mathbf{e_y})/\sqrt{2}$) as $\mathbf{a_1} = \sin(\varepsilon_1 - \pi/4)\mathbf{e_{-1}} + \cos(\varepsilon_1 - \pi/4)\mathbf{e_{+1}}$, and $\mathbf{a_2} = e^{-i\theta}\sin(\varepsilon_2 - \pi/4)\mathbf{e_{-1}} + e^{i\theta}\cos(\varepsilon_2 - \pi/4)\mathbf{e_{+1}}$, where $\varepsilon_j$ is the ellipticity parameter (angle) of the $j$-th component $-\pi/4 \leq \varepsilon_{1,2} \leq \pi/4$, $|\tan\varepsilon_j|$ equals the ratio of minor to major axis of the polarization ellipse (Fig.1a), and the sign of $\varepsilon_j$ governs the direction of the field vector rotation. $\theta$ is the angle between the major axes of the ellipses. We assume that atoms are in a magnetic field $\mathbf{B}$ directed along the $z$ axis (Fig.1a).

The field (1) drives the $\Lambda$-type two-photon resonance between the two ground-state hyperfine levels with total angular momenta $F_1$ and $F_2$ (where $F_1$ and $F_2$ may be either integers or half-integers). The wavefunctions of the Zeeman substates will be denoted as $|F_1, m\rangle$ and $|F_2, m\rangle$. The excited state with angular momentum $F_e$ (Fig.1a) and the corresponding Zeeman manifold is denoted as $\{|F_e, \mu\rangle\}$.

As was shown in the paper,[12] the general field configuration that generates pure superposition dark states, is $\varepsilon_1 \perp \varepsilon_2$. In this case the polarization ellipses of the components of bichromatic radiation are orthogonal ($\theta = \pi/2$). If the ellipticities obey the condition

$$\frac{1+F-m}{1+F+m} = \frac{\tan(\varepsilon_1 + \pi/4)}{\tan(\varepsilon_2 + \pi/4)}, \qquad (2)$$

there exists an atomic state, nullifying at exact two-photon resonance the Hamiltonian of interaction with bichromatic field. This state is a coherent superposition of the Zeeman substates $|F_1, m\rangle$ and $|F_2, m\rangle$ with the same magnetic quantum number $m$. If the Zeeman splitting is large enough, there are no other dark states. Consequently, in the course of optical pumping, the majority of atoms are accumulated in the given pure superposition $m$-$m$ state.

Among all possible $\varepsilon_1 \perp \varepsilon_2$ configurations we can distinguish a symmetric class $\varepsilon \perp (-\varepsilon)$ (i.e. $\varepsilon_1 = -\varepsilon_2$), where the condition (2) is reduced to

$$\sin(2\varepsilon) = -m/(1+F) \,. \tag{3}$$

Note that the $\varepsilon \perp (-\varepsilon)$ configuration can be obtained from the *lin* $\perp$ *lin* one simply by adding a quarter-wave plate into the beam path.

Besides of the analytic treatment of the problem, we performed numerical calculations as well. We used for this the quantum-kinetic equation for the atomic density matrix. We found that the population of the pure superposition state $|F_1, m\rangle - |F_2, m\rangle$ approximately equals the resonance contrast, which we define as a ratio of the resonance transmission to the Doppler absorption background.

To verify the above ideas, we detected the transmission of the $\varepsilon \perp (-\varepsilon)$ bichromatic field as a function of the two-photon detuning $(\omega_1 - \omega_2)$. To select the $|F_1, m\rangle - |F_2, m\rangle$ resonance, an axial magnetic field was applied to the atoms. When the pure superposition dark states are generated, each $m - m$ resonance has a pronounced maximum depending on the ellipticity $\varepsilon$ of applied two-photon field, (see eq. (3)).

The experimental setup is shown in Fig. 1a. We perform the experiment with atomic Rb vapor at 30 $C^0$, which corresponds to atomic density $N \sim 10^{10}\ cm^{-3}$. Our experiments use an atomic cell containing natural abundance of $^{87}$Rb and $^{85}$Rb. The length and diameter of the cell equals 7 *cm* and 2.5 *cm*, respectively. 3.0 *torr* of Ne was used as a buffer gas. The cell is made from pyrex glass and has windows with small birefringence. The cell is placed into a three layer $\mu$-metal anti-magnetic shield. A solenoid mounted inside the magnetic shield allows us to control the amplitude of the homogeneous magnetic field applied along the direction of the light propagation.

Two extended cavity diode lasers are used (Fig. 1c). The frequencies of the ECLD1 and ECLD2 are tuned to $|F_1\rangle \to |F_e\rangle$ and $|F_2\rangle \to |F_e\rangle$ transitions, respectively. We refer the transition $|F_2\rangle \to |F_e\rangle$ as $|5S_{1/2}, F = 2\rangle \leftrightarrow |5P_{1/2}, F' = 2\rangle$ in case of $^{87}Rb$ and $|5S_{1/2}, F = 3\rangle \leftrightarrow |5P_{1/2}, F' = 3\rangle$ - for $^{85}Rb$. Another leg of $\lambda$ transition $|F_1\rangle \to |F_e\rangle$ is referred to as $|5S_{1/2}, F = 1\rangle \leftrightarrow |5P_{1/2}, F' = 2\rangle$ in case of $^{87}Rb$ and $|5S_{1/2}, F = 2\rangle \leftrightarrow |5P_{1/2}, F' = 3\rangle$ - for $^{85}Rb$. The frequencies of the lasers are locked to each other by phase sensitive locking electronics. The locked laser frequency difference equals the ground state splitting of the isotope under investigation, (6.834 $GHz$($^{87}Rb$) or 3.035 $GHz$ ($^{85}Rb$)) and can be scanned over a 40 MHz range. All data were taken at $P_{ECLD1} = 1.68 mW$, $P_{ECLD2} = 0.32 mW$ powers, beam diameters equal *0.8mm* and *0.3mm* (Rabi frequencies - *(15-20) MHz*). Frequency separation between CPT resonances were almost half ( 6-10 MHz) the Rabi frequencies of the applied fields. As a result, the applied fields interact simultaneously with two (or three) Zeeman states, producing few pure superposition states. This gives an error in the estimation

of the number of dragged atoms contributing to the dark states ( for $Rb^{85}$ this error is larger compared to $Rb^{87}$, because the former has more Zeeman sublevels).

The orthogonally polarized laser beams, combined on the beamsplitter (PBS), enters into the cell. The polarization ellipticity of each beam is controlled by the $\lambda/4$ plate. The rotation of the $\lambda/4$ waveplate synchronously changes the ellipticity $\varepsilon$ of both laser beams with opposite sign.

The Doppler broadening of rubidium vapor (the full width at the half maximum is equal to $\sim 540\ MHz$) overlaps the hyperfine features of the $Rb^{85}\ ^2P_{1/2}$ state, whereas the hyperfine separation of the $Rb^{87}$ is larger and equals 812 $MHz$. These differences in isotopic structure are important to verify the vulnerability of our technique to hyperfine spectral resolution manifold.

To determine the population deposited into superposition states, we measured the contrast of the CPT resonance for different polarization ellipticities, see Fig. 2 and Fig. 3(left), respectively. As is seen, these dependencies have well pronounced extrema. Their positions correspond to eq. (3) (e.g. $\varepsilon = 0$ for $m = 0$ and $\varepsilon = \pm\pi/12$ for $m = \mp 1$). This is clear proof of the validity of theoretical treatment presented in.[12]

Our setup had slight uncontrollable ellipticity of polarization which is displayed in the asymmetry of the curves in the Fig. 2 and Fig. 3. The experimentally measured population of the $m$-$m$ states is in good agreement with our numerical calculation for $Rb^{87}$. In our experiments the number of atoms deposited into the $m-m$ states is limited by the intensities of our lasers. Increasing the laser intensity may increase the population of the $m-m$ states up to 100%.

In conclusion, we have proposed configurations of a bichromatic field, which allow the preparation (with the use of the CPT effect) of pure superpositions of the $m-m$ states for arbitrary magnetic quantum number. It has been shown that in the general case of $m \neq 0$ it is necessary to use elliptically polarized fields. We have found that if the hyperfine structure of the excited state is not resolved, the dark state superpositions in alkali atoms can be generated only in the D1 line. We deposited almost 60% of the atoms into the 0-0 pure state of $Rb^{87}$. These results may have applications in atomic clocks, magnetometers, quantum optics, and atom cooling.

We thank L. Hollberg, H. Robinson, J. Kitching, S. Knappe, V. Shah, and Y.-Y. Jau for helpful discussions. This work was supported by RFBR (grants 05-02-17086, 04-02-16488, and 05-08-01389) and by a grant INTAS-01-0855. VLV and SAZ were supported by a grant ISTC 2651p.

## List of Figures

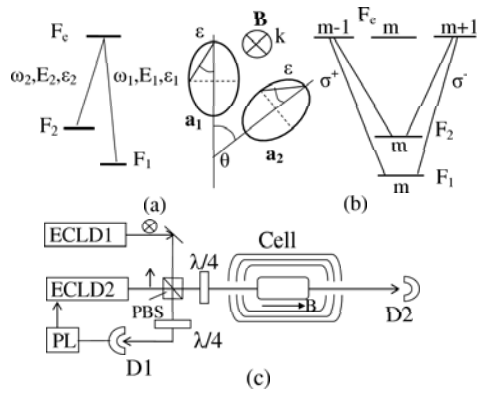

Fig. 1. a) Three-level $\Lambda$ scheme, polarization ellipses $\mathbf{a_1}$, $\mathbf{a_2}$ and orientation of the magnetic field $\mathbf{B}$, wave vector ($\mathbf{k}$) applied in the experiment. b) Light-induced transitions driven by the $\sigma^+$ and $\sigma^-$ circularly polarized components for the $m-m$ resonance; c) Experimental setup. ECLD - extended cavity laser diodes, PL - phase locking electronics, D1,D2- photodiodes.

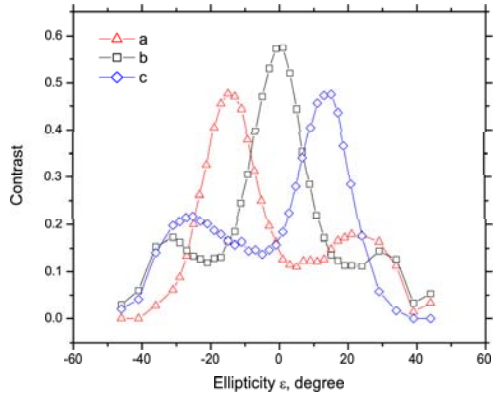

Fig. 2. Contrast of the $Rb^{87}$ CPT resonances on the ellipticity $\varepsilon$ of the polarization at different transitions: a) $|F=1, m_F=-1\rangle \leftrightarrow |F=1, m_F=-1\rangle$; b) $|F=1, m_F=0\rangle \leftrightarrow |F=1, m_F=0\rangle$; c) $|F=1, m_F=+1\rangle \leftrightarrow |F=1, m_F=+1\rangle$. Accumulated population in the state reaches *51%, 58%, 51%*, correspondingly.

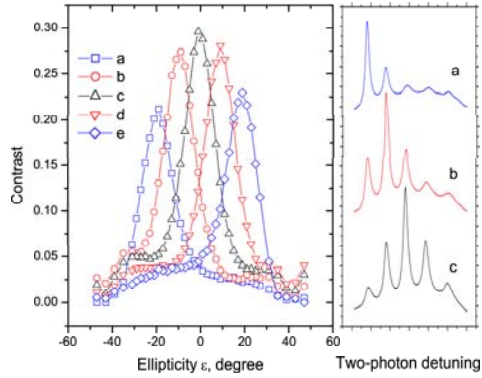

Fig. 3. (left) Contrast of the $Rb^{85}$ CPT resonances, on the ellipticity $\varepsilon$ of the polarization at different transitions: a) $|F=2, m=-2\rangle \leftrightarrow |F=3, m=-2\rangle$; b) $|F=2, m=-1\rangle \leftrightarrow |F=3, m=-1\rangle$; c) $|F=2, m=0\rangle \leftrightarrow |F=3, m=0\rangle$; d) $|F=2, m=+1\rangle \leftrightarrow |F=3, m=+1\rangle$; e) $|F=2, m=+2\rangle \leftrightarrow |F=3, m=+2\rangle$. Accumulated population in states reaches 25%, 30%, 31%, 30%, 25%, respectively.
(right) Transmission of the $Rb^{85}$ CPT resonances on the two-photon detuning at different polarization ellipticity ($\omega_1$ is swept and $\omega_2$ is tuned to the maximum of Doppler absorption ). Full frequency scanning range equals 40 MHz. a) $\varepsilon=-21^0$, b) $\varepsilon=-9^0$ c) $\varepsilon=0^0$ ellipticity of the applied polarization components.